\documentstyle[twocolumn,aps,prl,eqsecnum]{revtex}
\begin{document}

\title{Breaking of vortex lines - a new mechanism of collapse 
 in hydrodynamics}
\author{E.A.Kuznetsov \cite{1} and V.P.Ruban \cite{2}}
\address{\it L.D.Landau Institute for Theoretical Physics,
2 Kosygin str., 117334 Moscow, Russia. }


\maketitle

\small

\begin{abstract}
A new mechanism of the collapse in hydrodynamics is suggested, due to
breaking of continuously distributed vortex lines. Collapse results in
formation of the point singularities of the vorticity field $|{\bf\Omega}|$.
At the collapse point, the value of the vorticity blows up as $(t_0-t)^{-1}$ 
where $t_0$ is a collapse time. The spatial structure of the collapsing 
distribution approaches a pancake form: contraction occurs by the law 
$l_1\sim(t_0-t)^{3/2}$ along the "soft" direction, the characteristic scales 
vanish like $l_2\sim(t_0-t)^{1/2}$ along two other ("hard") directions. 
This scenario of the collapse is shown to take place in the integrable 
three-dimensional hydrodynamics with the Hamiltonian 
${\cal H}=\int|{\bf\Omega}|d{\bf r}$. Most numerical studies of collapse 
in the Euler equation are in a good agreement with the proposed theory.
\end{abstract}




\section{Introduction}

Collapse in hydrodynamics of an ideal incompressible fluid, as a process of
singularity formation in a finite time, is one of the central problems in
the theory of developed hydrodynamic turbulence.
The classical examples of such type spectra are
the Phillips spectrum for water-wind waves \cite{Phillips} and the Kadomtsev-Petviashvili spectrum
for acoustic turbulence \cite{KP}.  In the first case white caps -- wages of water surface
-- play a role of singularities, and in the second case these are density breaks (or
shocks).

The question about collapse in hydrodynamics is an old problem.
For example, in 1981 P.Saffman \cite{saffman} considered collapse as
one of the most important problems in hydrodynamics
(see also papers \cite{PS} and references therein),
probably already L.Richardson and A.N.Kolmogorov
understood an importance of this problem.
In spite of so long a history of the question, there is no deep
understanding of the nature of a collapse in hydrodynamics, 
even though there are many numerical
simulations testifying to the collapse existence collected by now.
As far as theory is concerned, essential results are absent, and, moreover,
there is no common agreement about collapse as a subject for incompressible
hydrodynamics (see, e.g., Sec. 7.8 of the book of U.Frisch \cite{frisch}
and references therein).
 In the theory, the only appreciable exclusion is the
work of V.E.Zakharov, 1988 \cite{zakh} (more detailed publication has
appeared  in  1999 \cite{zakh99}), where the consistent theory of
collapse for two anti-parallel vortex filaments of small thickness was
developed in quasi-two-dimensional approximation, when a flow is almost
two-dimensional with a slow  dependence with respect to the third coordinate
(see also the paper \cite{KMD}).
A significant progress in studying hydrodynamic collapse has been achieved in
numerical simulations  of the Euler equation. Many numerical experiments
testify that the value of the vorticity $|{\bf \Omega}|$ becomes infinite
in isolated points in a finite time. As follows from the papers of
Kerr \cite{kerr}, Grauer, Marliani and Germashewsky \cite{GMG}, Pelz
\cite{P}, Boratav and Pelz \cite{BP}, $|{\bf \Omega}|$ grows at the
collapse point like $(t_0-t)^{-1}$, where $t_0$ is a collapse time.
According to \cite{kerr}, \cite{BP}, a spatial scale of the collapsing
distribution contracts as $(t_0-t)^{1/2}$. In the recent paper by Kerr
\cite{kerr1}, an anisotropy of the collapsing region has been reported.
The data processing gave two scales, one of them being contracted as the root
$l_1\sim (t_0-t)^{1/2}$, and another one as $l_2\sim t_0-t$. It should be
noted that the initial flow either possessed a
definite symmetry or it was close to a symmetric 
flow in most numerical simulations. As a result,
several singularities arose simultaneously. 
For example, the evolution of two anti-parallel vortex tubes was studied in
\cite{kerr}. The collapse here is caused by the Crow instability \cite{crow}
leading at the nonlinear stage to the vortex reconnection.  
Symmetry of the flow makes the
collapse to happen in two symmetric points.

In the present paper, we suggest a new mechanism of the singularity
formation connected with breaking of continuously distributed
vortex lines. This mechanism is not related to any
symmetry of the initial vorticity.
The collapse itself is possible in one separate point.
Probably, just this type of collapse has been observed in the
recent numerical experiment \cite{sinai}.

The mechanism suggested can be naturally incorporated into the
classical catastrophe theory \cite{arnold}. From this point of view,
collapse can be considered as caustic formation for a solenoidal field.
It is not so easy to understand, how a collapse arises in the Eulerian
description. First, this is connected with a hidden symmetry of the Euler
equation, i.e., the relabeling symmetry (for more details see 
the reviews \cite{salmon}, \cite{ZK}). This symmetry generates the 
conservation law for the Lagrangian invariant -- the so-called  Cauchy 
invariant, which is expressed through the velocity curl and the Jacoby matrix 
of a mapping from the Eulerian variables to the Lagrangian ones, 
and by this reason this invariant
occurs very nonlocal in terms of the velocity field. On the one hand,
the Cauchy invariant is known as invariant which characterizes the property
of frozenness of  vortex lines into a fluid. On the other hand, all
known conservation laws for vorticity, such as the Kelvin's and
Ertel's theorems, conservation of the topological Hopf invariant, 
being a measure of the flow
knottiness, are a simple consequence of the Cauchy invariant constancy. The
frozenness of vortex lines means that fluid particles are pasted to
a given vortex line and never leave it. A destruction of frozenness is
possible only due to viscosity, i.e., beyond ideal
hydrodynamics. Therefore, as a next natural step in the vortex motion
description, a mixed Lagrangian-Eulerian description has been introduced,
where the main object is a vortex line \cite{KR},\cite{KR00}. Each vortex
line in this description is labeled by a two-dimensional Lagrangian marker, 
while the third coordinate serves as a parameter determining the curve. 
This representation, which we called as the vortex line representation, 
is a key-point in the description of  hydrodynamic
collapse which can be considered as a process of a caustic 
formation for the solenoidal vorticity field.

The paper is organized as follows: in Sec.II we introduce the vortex line
representation  and explain its meaning. In Sec.III
we consider the 3D integrable hydrodynamic model introduced in our previous
paper \cite{KR}.
The Hamiltonian of this model is unusual, it is expressed through the
absolute value of the vorticity $|{\bf\Omega}|$
\begin{equation}
\label{H}
{\cal H}=\int |{\bf \Omega }|d{\bf r}.
\end{equation}
The given model can be integrated by means of combination of the
vortex line representation and the inverse scattering transform.
By applying the vortex line representation, the Hamiltonian is 
decomposed into a sum of Hamiltonians of non-interacting vortex lines. 
Dynamics of each vortex line is described by the integrable one-dimensional 
Landau-Lifshitz equation for a Heisenberg ferromagnet or by its 
gauge-equivalent - the nonlinear Schroedinger equation. Thus, the integrable
hydrodynamics represents a hydrodynamics of free vortex lines. As
for hydrodynamics of free particles -- hydrodynamics of dust with a
null pressure (see, e.g. \cite{zeld}), for the 3D integrable
hydrodynamics typical singularities are also caustics. For hydrodynamics
of dust, density  turns into infinity at caustics. Unlike
the dust density, being a scalar characteristics, vorticity is
a vector solinoidal field. Therefore, the latter imposes some
restrictions on a spatial structure near singularity.  As it
is shown in Sec.IV, the singularity structure, being very
anisotropic, turns into a pancake form. The spatial collapsing distribution
at $t\rightarrow t_0$ leads to quasi-two-dimensional. 
Along the "soft" direction a more rapid
compression takes place as $l_1\sim (t_0-t)^{3/2}$, and along two
other-- "hard" -- directions  $l_2\sim (t_0-t)^{1/2}$. At
the collapse time, the vorticity vector lies in the pancake plane and its
value $|{\bf \Omega}|$ blows up like $(t_0-t)^{-1}$. This behavior
corresponds to the general situation. The degenerated case is considered
in Sect. V, where we consider   collapse for
topologically nontrivial axi-symmetric distribution of
vorticity  in the form of the so-called Hopf mapping when any two
vortex lines are linked once with each other. In this case  two eigen-values
of the Jacobi matrix vanish
simultaneously  in the collapse point. As its sequence,  vorticity
occurs to have more strong singularity:
$|{\bf \Omega}|\sim (t_0-t)^{-2}$. And we conclude in Sec. VI with
discussion of numerical experiments on collapse observation in the
Euler equation and their correspondence to the proposed theory.

\section{Vortex lines representation of hydrodynamics}

Let us consider the equations of the hydrodynamic type
\begin{equation}
\label{dOmega_dt}
\frac{\partial {\bf \Omega}}{\partial t}=\mbox{curl}
\left[\mbox{curl}\frac{\delta {\cal H}}{\delta {\bf \Omega}}\times
{\bf \Omega}\right],
\end{equation}
where ${\cal H}\{{\bf \Omega}\}$ is the Hamiltonian of a system,
${\bf \Omega}({\bf r},t)=\mbox{curl\,}{\bf p}({\bf r},t)$ is 
the generalized vorticity, ${\bf p}$ is the canonical momentum. 
The vector field
\begin{equation}
\label{vel}
 {\bf v}=\mbox{curl}({\delta {\cal H}}/{\delta {\bf
\Omega}})
\end{equation}
is nothing else but the fluid velocity. By the definition
$\mbox{div\, }{\bf v}=0$, i.e., we deal with an incompressible fluid.
If the Hamiltonian coincides with the kinetic energy of the fluid
$$
{\cal H}=\int \frac{ {\bf p}^2}{2} d{\bf r}=
\int\!\int
\frac{{\bf\Omega}({\bf r}_1)\cdot{\bf\Omega}({\bf r}_2)}
{8\pi|{\bf r}_1-{\bf r}_2|}d{\bf r}_1d{\bf r}_2
$$
then the expression (\ref{vel}) yields the usual relation
${\bf \Omega}=\mbox{curl}\,{\bf v}$ between  velocity ${\bf v}$ and
vorticity ${\bf \Omega}$, while Eq.(\ref{dOmega_dt}) transforms
into the Euler equation for vorticity
$$
\frac{\partial{\bf \Omega}}{\partial t}=
\mbox{curl}[{\bf v\times\Omega}], \qquad \mbox{div}\,{\bf v}=0.
$$
The important property of the equation (\ref{dOmega_dt})
is the frozenness of vorticity into the substance,
i.e. all Lagrangian fluid particles at $t>0$ remain at their own vortex line.
After that, it is natural to introduce a mixed Lagrangian-Eulerian
description when each vortex line is labeled by its own
Lagrangian marker ${\bf \nu}$, which lies in a fixed two-dimensional manifold
${\cal N}$, while a parameter $s$ along the line has a meaning of an
Eulerian variable. In such vortex line representation vorticity is
expressed as follows \cite{KR}
\begin{equation}
\label{rh}
{\bf \Omega}({\bf r},t)=\int_{\cal N} \Omega_0(\nu)d^2\nu \int
\delta({\bf r}-{\bf R}(s,\nu,t))
\frac{\partial\bf R}{\partial s} ds.
\end{equation}
Here, the closed curve ${\bf r}={\bf R}(s,\nu,t)$ corresponds
to each vortex line $\nu$, so that ${\bf R}_s$ is its tangent vector.
The quantity $\Omega_0(\nu)$,  with ${\bf R}_s$ being its tangent vector.
The fixed function $\Omega_0(\nu)$ is the strength of vortex loop.
However, without loss of generality, this
function can be put equal to the unity. This can be achieved by both an 
appropriate
re-definition of labels $\nu$ and changing the vortex
orientation to the opposite one for those lines from the manifold  ${\cal N}$,
for which $\Omega_0(\nu)<0$. Therefore, in the next
section we will omit the multiplier $\Omega_0(\nu)$ in front of $d^2\nu$ in
the corresponding formulae.

The generalization of Eq. (\ref{rh})
to the case of arbitrary topology of the vortex lines is served by
the formula:
\begin{equation} \label{OmegaR}
{\bf \Omega}({\bf r},t)=\int\delta({\bf r}-{\bf R}({\bf a},t))
({\bf \Omega}_0({\bf a})\nabla_{\bf a}){\bf R}({\bf a},t)d^3{\bf a},
\end{equation}
where ${\bf \Omega_0({\bf a})}$ is the Cauchy invariant,
characterizing the frozenness property. The $\mbox{div}_{\bf
a}\,{\bf \Omega}_0({\bf a})=0$ condition guarantees automatically
incompressibility for the field ${\bf\Omega(r},t)$:
$\mbox{div}\,{\bf \Omega}({\bf r})=0$. In the expression
(\ref{OmegaR}), the vector 
$$
{\bf b}=({\bf\Omega}_0({\bf a})\nabla_a){\bf R}({\bf a},t)
$$ 
is a tangent vector to the vortex line at the
point
\begin{equation}
\label{map}
{\bf r}={\bf R}({\bf a},t).
\end{equation}
In the representation  (\ref{rh}), an arc-length of line
of the initial field ${\bf \Omega}_0({\bf a})$ can serve as the parameter $s$.

After integrating  (\ref{OmegaR})  over ${\bf a}$-variables, the vector
${\bf \Omega(r},t)$ is expressed through the Jacobian $J$ of the mapping
(\ref{map}) $J={\mbox{det}||\partial{\bf R}/\partial {\bf a}||}$ and the
Cauchy invariant ${\bf \Omega_0({\bf a})}$:
\begin{equation}
\label{O_det}
{\bf \Omega}({\bf R})=
\frac 1J{({\bf \Omega}_0({\bf a})\nabla_{\bf a}){\bf R}({\bf a})}.
\end{equation}
It is important to emphasize, that in this expression the Jacobian is not
to be equal to the unity: $J\neq 1$. Nevertheless, it does not
contradict to the incompressibility condition for the fluid.

As it was shown in  \cite{KR,KR00}, the equations of motion for vortex
lines can be obtained directly from the equation of frozenness (\ref{dOmega_dt})
\begin{equation}
\label{R}
[\{({\bf\Omega}_0({\bf a})\nabla_ a){\bf R}({\bf a},t)\}\times
\{{\bf R}_t({\bf a},t)- {\bf v(R(a},t),t)\}]=0.
\end{equation}
This equation describes the {\it transverse} dynamics of vortex
line: obviously any motion along a curve leaves the curve unchanged.
In particular, this helps to understand why there are no
restrictions imposed on the value of the Jacobian $J$. Let us recall that in
a purely Lagrangian description of an incompressible fluid Jacobian
is equal to the unity identically. But accordingly to the equation
(\ref{R}), the motion of Lagrangian particles along vortex
lines is excluded from the mapping (\ref{map}) for the mixed 
Lagrangian-Eulerian description. It is exactly the reason why
the Jacobian is not necessarily equal to unity now. This point is
principal and will be instrumental below in explaining how collapse
is possible in the hydrodynamic systems (\ref{dOmega_dt}).

 As it was shown in  \cite{KR,KR00}, Eq.(\ref{R}) can be written in the
Hamiltonian form
\begin{equation}
\label{main}
\left[\{({\bf\Omega}_0({\bf a})\nabla_ a){\bf R}({\bf a},t)\}
\times{\bf R}_t({\bf a})
\right]=
\frac{\delta {\cal H}
\{{\bf \Omega}\{{\bf R}\}\}}
{\delta{\bf R}({\bf a})}\Big|_{{\bf \Omega}_0}.
\end{equation}
This equation  describes a motion of vortex
lines in systems with an arbitrary
Hamiltonian that depends on ${\bf R}$ through the
${\bf\Omega(r},t)$ only.

It is useful also to keep in mind that the expressions for such important
characteristics of the system as its momentum ${\bf P}=\int{\bf p}\,d{\bf r}$
and  angular momentum ${\bf M}=\int[{\bf r}\times{\bf p}]d{\bf r}$,
being transformed by integration by parts to a form, where, instead of
${\bf p}$, the vorticity ${\bf\Omega}$ is employed, and being then rewritten
in terms of vortex lines, have the form
\begin{equation}\label{mom}
{\bf P}\sim\frac{1}{2}\int{\bf [r\times\Omega]}d{\bf r}=
\int_{\cal N}d^2\nu\,\,\frac{1}{2}\int[{\bf R}\times{\bf R}_s]ds
\end{equation}
\begin{equation}\label{anglmom}
{\bf M}\sim \frac {1}{3}\int {\bf [r\times [r\times \Omega}]]d{\bf r}
=\int_{\cal N}d^2\nu\,\,\frac {1}{3}\int {\bf [R\times [R\times R}_s]]ds
\end{equation}
The $\sim$-sign in these relations means that equalities take place up to
integrals over surface with infinitely large radius.  
Hence one can see that the
momentum and the angular momentum are composed of momenta and angular momenta
of each vortex line, the momentum of a closed line being equal to the oriented
area of a surface tightened on  the vortex loop.

It is easily to verify that uniform shift ${\bf R_0}$ of ${\bf R}$
does not change the momentum, while the angular momentum is
subjected to the well known transformation
\begin{equation}
\label{ang}
{\bf M} \rightarrow {\bf M'}= {\bf [R_0\times P]} + {\bf M}.
\end{equation}

\section{Integrable hydrodynamics}

In this and two next sections,
we will show how and why  collapse is possible in 3D
integrable hydrodynamics. This model was introduced in our previous paper
\cite{KR}. The Hamiltonian of this model is expressed through the absolute 
value of ${\bf \Omega(r},t)$
\begin{equation}\label{Hsimplest}
{\cal H}=\int |{\bf \Omega(r) }|d{\bf r},  \label{hamilt}
\end{equation}
and the equation of motion coincides with the
frozenness equation (\ref{dOmega_dt}) with velocity
\[
{\bf v}=\mbox{curl\,}{\vec \tau}
\]
where ${\vec\tau}=\left( {\bf \Omega }/{\Omega }\right)$
is the unit tangent vector along the vortex line. Assuming all the
lines  closed, choosing the labeling by such a way so
that  $\Omega_0(\nu)=1$, and
substituting the representation (\ref{rh}) into (\ref{hamilt}),
it is easy to see that the
Hamiltonian is decomposed as a sum of Hamiltonians for the vortex
lines \footnote{It is worth to notice that this property is common for all systems with the
Hamiltonians of the type ${\cal H}=\int
F(\tau,(\tau\nabla)\tau, (\tau\nabla)^2\tau,\dots)|{\bf \Omega }|d{\bf r}$.
To explain  the idea of  collapse of vortex lines, we have
chosen the simplest example (\ref{Hsimplest}),
which has a physical meaning.}:
\begin{equation}
\label{ham}
{\cal H}\{{\bf R}\}=\int d^2\nu \int
\left|\frac{\partial\bf R}{\partial s}\right| ds.
\end{equation}
Here, the integral over $s$
is the length of the vortex line with the index $\nu$.
The equation of motion for the vector ${\bf R}(\nu,s)$,
in accordance with the Eq. (\ref{main}), is
local in these variables  -- it doesn't contain an interaction with other
vortices:
\begin{equation}
\label{motion}
 [{\bf R}_s\times{\bf R}_t]=
 [{\vec\tau}\times[{\vec\tau}\times{\vec\tau}_s]].
\end{equation}
By this reason, not only the total energy, momentum,
and angular momentum
are conserved, but also the corresponding geometrical invariants for each
vortex loop: its length 
$$
{\cal H}(\nu)=\int |{\bf R}_s(\nu)|ds ,
$$
the oriented area spanned on the vortex loop which  coincides with 
its momentumits momentum
$$
 {\bf P}(\nu)= \frac 12 \int [{\bf R}(\nu)\times {\bf R}_s(\nu)]ds,
$$
and its angular momentum
$$
{\bf M}(\nu)=
\frac 13 \int [{\bf R}(\nu)\times [{\bf R}(\nu)\times {\bf R}_s(\nu)]] ds .
$$

It is important to pay attention to the following fact: 
The equation (\ref{motion}) is invariant with respect to changes
$s\rightarrow \tilde s(s,t)$.
Therefore, it can be solved for  ${\bf R}_t$
up to a shift along the vortex line  -- the transformation leaves 
the vorticity
${\bf \Omega}$ unchanged.
This means that to find the vorticity ${\bf \Omega}$ it is enough to have one
solution of the  equation
\begin{equation}
\label{motion1}
|{\bf R}_s| {\bf R}_t=
[{\vec\tau}\times{\vec\tau}_s] + \beta{\bf R}_s
\end{equation}
which
follows from Eq.(\ref{motion}) for some choice $\beta$.
This leads to an equation for ${\vec \tau}$ as a function of the
filament length $l$ ($dl=|{\bf R}_s|ds$) and time $t$ (by choosing a new value
$\beta=0$), which reduces to the integrable  one-dimensional (1D)
Landau-Lifshits
equation  for a Heisenberg ferromagnet
\begin{equation}
\label{fer}
\frac{\partial {\vec \tau}}{\partial t}=
\left [{\vec\tau}\times\frac{\partial ^2{\vec\tau}}{\partial l^2}
\right].
\end{equation}
This equation, in its turn, is gauge equivalent to the 1D nonlinear
Shr{\"o}dinger equation \cite{ZT}
\begin{equation}
\label{schr}
i\psi_t+\psi_{ll}+(1/2)|\psi|^2\psi=0
\end{equation}
and, for instance, can be reduced to the NLSE  by means of the Hasimoto
transformation \cite{Hasimoto}
$$
\psi(l,t)=\kappa(l,t)\cdot
\exp\left[i\int^l \chi(\tilde l,t)d\tilde l\right],
$$
where $\kappa(l,t)$ is a curvature and $\chi(l,t)$
the line torsion.

The system under consideration has  direct relation to hydrodynamics.
As it is known \cite{rios},\cite{betchov},
the local induction approximation for a thin vortex
filament, under assumption of smallness of  the filament width
to the characteristic longitudinal scale, leads to the Hamiltonian
(\ref{ham}), but only for a single separate line. 
The essence of this approximation is in in replacing the
logarithmic interaction law by a delta-functional one.
When the widths of the filaments are
small comparable with distances between them, in the same approximation, 
the Hamiltonian of vortex lines transforms into the sum of the
Hamiltonians of independent vortex loops, yielding in a "continuous" limit
the Hamiltonian (\ref{hamilt}).

By such a way, we have the model of 3D integrable hydrodynamics of free
vortex filaments. In this model, each vortex is a nonlinear object with its
own internal dynamics. As we will see later, already in the framework of 
this model,
a singularity formation  is possible. Singularities appear in this model as a
result of intersection of vortex lines that is  analogous to the
phenomenon of wave breaking in gas-dynamics.

\subsection{Stationary vortices}

Let us consider now the simplest solution of Eq.(\ref{motion}),
i.e., a stationary propagation of a closed vortex line:
${\bf R}_t ={\bf V}\equiv \mbox{const}$.
In this case the velocity ${\bf V}$ is determined from
solution of the equation
\begin{equation}
\label{stat}
[{\bf R}_s
\times{\bf V}]=
[{\vec\tau}\times[{\vec\tau}\times{\vec\tau}_s]].
\end{equation}
It is easily to check that this equation follows from the variational
principle
\begin{equation}
\label{var}
\delta({\cal H}(\nu)-{\bf V\cdot P}(\nu))=0,
\end{equation}
i.e., any solution of (\ref{stat})
represents a stationary point of the Hamiltonian for a fixed momentum ${\bf
P}(\nu)$. The equation (\ref{stat}) can be simply integrated,
being rewritten in terms of the
binormal ${\bf b}$ and  the curvature $\kappa$ of the line as follows
\begin{equation}
\label{stat1}
[{\mathbf \tau}
\times{\bf V}]=
\kappa{\mathbf [\tau\times b]},
\end{equation}
that gives
\begin{equation}
\label{statsol}
{\bf V}= \kappa{\bf b}.
\end{equation}

A constant value of the velocity ${\bf V}$
in this expression implies constancy
 of the curvature $\kappa$, i.e. the vortex line must be a ring of
radius $r=1/\kappa$ and
\begin{equation}
\label{statsol1}
V= 1/r.
\end{equation}
The direction of the ring motion is perpendicular to its plane. It is
interesting to note that the  exact answer to the velocity of a thin
 (with width $d\ll r$) vortex ring in ideal hydrodynamics (\cite{lamb})
coincides with Eq.(\ref{statsol}) up to the logarithmic accuracy that
just differs the considering model from the Euler equation.

Stationary solutions (\ref{statsol})
in the form of rings are remarkable within this model, because
they are stable, moreover, they are stable in the Lyapunov' sense.
Remind, that momentum ${\bf P}$ of a closed vortex line is  its oriented
surface spanned on the loop:
$$
{\bf P}= S{\bf n},
$$
where $S$ is the surface value, ${\bf n}$ its normal. Inasmuch
as the Hamiltonian of a vortex loop  coincides with its length,
a maximum of the momentum, or, that is the same as a maximum of
surface $S$ is  obviously attained, for fixed length, at the perfect
circle. Just this proves stability of the vortex ring solution (\ref{statsol})
in the Lyapunov sense.

\section {Collapse}

The solution  (\ref{statsol}), (\ref{statsol1})
enables us to construct the simplest mappings  ${\bf R=R}(\nu,s,t)$.

Let all vortex lines be circle-shaped and oriented in the same direction,
for instance, along $z$-axis. We will see further that collapse in our
model is a purely local phenomenon. Therefore,
it is sufficient to consider some vortex tube
(which can be imagined as a torus) to find a mapping.
Let vortex rings be distributed continuously inside the tube. 
We label each vortex
line by the two-dimensional
parameter $\nu$, which values coincide with coordinate of some cross section of the  tube
at $t=0$. We will use the ring arc-length as  longitudinal parameter $s$
($ds=r d\phi$, where $\phi$ is the polar angle around $z$-axis). 
Then, with the help of
(\ref{statsol}), the desired mapping can be written  as follows
\begin{equation}
\label{solutionR}
{\bf R}= {\bf R_0}(\nu) +r(\nu)\mbox{cos}\,\phi\,{\bf e}_x+
r(\nu)\mbox{sin}\,\phi\,{\bf e}_y+V(\nu)t\,{\bf e}_z.
\end{equation}
In this formula ${\bf e}_{x,y,z}$
are unit vectors along the corresponding axes.

It can be easily verified for this mapping that the Jacobian is a linear
function of time
\begin{equation}
\label{jac}
J=\frac{\partial(X,Y,Z)}{\partial(\nu_1,\nu_2,s))}=
J_0(\nu,s )+A(\nu,s)t.
\end{equation}
Here $A(\nu, s)$ is a coefficient linearly dependent on the
velocity derivatives with respect to $\nu$ and $J_0$ the initial
value of Jacobian.

Dependence $J$ (\ref{jac}) on time means that for every fixed pair
$\nu$ and $s$ there exists
such a moment of time $t=\tilde t(\nu,s)$ ( $t>0$, or $t<0$),
when  Jacobian is equal to zero: $J(\nu,s,t)=0.$
Denote as $t_0$ the minimal value of $t=\tilde t(\nu,s)$
at $t>0$. And let this minimum be attained at some point ${\bf a}={\bf a_0}$
(here we denote a point $(\nu_1,\nu_2,s)$ as  ${\bf a}$).
It is evident that at $t=t_0$
$
\left(\partial{\tilde t}/\partial {\bf a}\right)|_{a=a_0}=0
$
or
\begin{equation}
\label{min}
\left.\nabla_aJ({\bf a},t)\right|_{a=a_0}=0,
\end{equation}
since
$$
\left.\nabla_aJ({\bf a},t)\right|_{a=a_0} +
\left.\frac{\partial J({\bf a},t)}{\partial t}
\frac{\partial{\tilde t}}{\partial {\bf a}}\right|_{a=a_0}=0.
$$
It is clear also that at $t=t_0$ the tensor of second
derivatives of $J$ against ${\bf a}$,
$$
2\gamma_{ij}=\frac{\partial^2 J }{\partial a_i\partial a_j},
$$
will be positive definite at the point ${\bf a= a_0}$. Hence,
it is easy to define a behavior of the Jacobian in a small
vicinity of ${\bf a=a_0}$. Expansion of $J$
near this point (in a typical situation) at $t\rightarrow t_0$ is as follows
\begin{equation}
\label{expansion}
J({\bf a},t)=\alpha(t_0-t)+\gamma_{ij}\Delta a_i\Delta a_j + ...,
\end{equation}
where
$$
\alpha=-\left.\frac{\partial J({\bf a},t)}{\partial t}\right|_{t=t_0,
a=a_0}>0,\,\,\, \Delta {\bf a=a-a_0}.
$$
Those are  the leading contributions to the Jacobian expansion
\footnote{
For instance, the term containing mixed time-space derivatives, in the general
case, is a small correction to the first term in (\ref{expansion}).}.

Geometrically, the above expansion  corresponds to a sufficiently simple
picture. The (hyper-) surface $J=J({\bf a},t)$
deforms with time in such a manner, that its minimum reaches the
(hyper-) plane  $J=0$ at $t=t_0$, where two surfaces touch each
other. Obviously, for smooth mappings in a typical case, this touching
takes place in one separate point. In a degenerated situation,  touching is
possible in a few points simultaneously, or even at a curve. A case, when
two eigenvalues of the Jacoby matrix $\hat J$ tend to zero simultaneously
at the collapse point, will be regarded  also as degenerated (such an example 
will be
considered in the next section). In that case one should keep the next
terms in the Jacobian expansion along the corresponding directions.
We would like to repeat that all these cases can not be considered as  
typical ones.

In accordance with Eq. (\ref{O_det}), the equality $J=0$
at the singular point means the formation of a singularity for the
vorticity at the moment $t=t_0$:
\begin{equation}
\label{omega}
{\bf \Omega(r},t)=\frac{\Omega_0(\nu){\bf R}_s}{\alpha(t_0-t)+
 \gamma_{ij}\Delta a_i\Delta a_j }.
\end{equation}
It is important, that the numerator in this fraction --
the tangent vector  of a vortex line -- does not vanish at the point
${\bf a}_0$ due  to its geometrical meaning.
Therefore, the vorticity at the singular point blows up as
$(t_0-t)^{-1}$, and the characteristic size of the collapsing distribution 
in $a$-coordinates decreases
 as $\sqrt{t_0-t}$.

The above type of collapse arises as a result of vortex line
breaking when one vortex  overtakes another. For
flows of the general type (without symmetries) a singularity must arise
at the first time always in  one separate point.

As we will see further, {\it for the given type of collapse},
the dependence (\ref{omega}) for  ${\bf \Omega(r},t)$,
derived for the particular initial distribution, is actually the general
answer, which can be applied not only to the integrable hydrodynamics
but also to the whole family (\ref{dOmega_dt}) of hydrodynamic systems,
of course, under the condition that they
admit such (quasi-) inertial regime of collapse.
What necessary conditions should satisfy the
Hamiltonian of some particular system, in order to exhibit such regime?
Now the answer to this question has been unknown that provides a wide field for future
investigations, both theoretical and numerical.

\subsection{Non-stationary vortices}

Let us consider now the integrable hydrodynamics in a more general case when
closed vortex lines are not circles. In this situation, it is possible 
to map each vortex contour to some vortex ring. It is then  natural to
introduce the (mean) direction ${\bf n}$, as well as the mean area
$S=\pi r_0^2$, with the help of the expression for momentum of vortex line,
${\bf P=n}S.$
The position ${\bf R_0}$  of the ring center
changes linearly in time. A corresponding
mean velocity of the motion of closed line must be directed along the
momentum, in order to satisfy the conservation law for angular momentum.
The mean velocity value ${\bf V_0}$, generally speaking, is a function of those fundamental
integrals of motion, which are independent on the origin choice --
the Hamiltonian (i.e. the length $L$), the momentum, and the projection
on the vector ${\bf n}$ of the angular momentum. 
It is clear also that increasing the contour sizes in $\lambda$ times
should result in decreasing the velocity in $\lambda$ times, so that
\begin{equation}
\label{meanvel}
{\bf V_0}= \frac{8}{\pi^2} \frac{{\bf P}}{L^3}\cdot
U\left(\frac{16\pi^2 P^2}{L^4},
\frac{({\bf M}\cdot{\bf P})}
{L^5}\right).
\end{equation}
Explicit dependence of the $U(\xi,\eta)$-function has been unknown
now. Its first argument can be considered as a measure of a "crumpleness" of the vortex
line and  changes in the limits  $0\le\xi\le1$, and the second argument
determines a measure of a spirality of the line.
In a more or less reasonable approximation, which doesn't lead
to excessive errors, one can suppose that $U(\xi,\eta)\sim U(1,0)=1$.

After introducing the mean characteristics, the mapping ${\bf R}(\nu,s,t)$
can be represented as the sum:
\begin{equation}
\label{mean}
{\bf R(\nu},s,t) = \tilde {\bf R}(\nu,s,t) + \delta {\bf r}(\nu,s,t),
\end{equation}
where the mean value $ \tilde {\bf R}(\nu,s,t)$   is given by the relation
\begin{equation}
\label{mean1}
{\bf \tilde R}(\nu,s,t) = {\bf R_0}  +r_0\mbox{cos}\,\phi'\,{\bf e'}_x+
r_0\mbox{sin}\,\phi'\,{\bf e'}_y,\qquad
\dot{\bf R}_0=V_0{\bf n}.
\end{equation}
Here  angular parameter $\phi'=2\pi(s/L)$ is proportional to the
arc-length $s$, and unit vectors ${\bf e'}_x ,{\bf e'}_y$ lie
in the plane, perpendicular to the local
$z'$-axis, directed along  the ${\bf n}$.
 The relationship between  $r_0$
and $V_0$ is given by the Eq. (\ref{meanvel}).
The vector function $\delta {\bf r(\nu,}s,t)$ describes deviations
(generally speaking, not small) from the mean value $\tilde{\bf R}(\nu,s,t)$.

The separation of the mean and oscillatory motions, introduced by means of
 (\ref{mean}), (\ref{mean1}), (\ref{meanvel}) for each vortex contour,
shows that the mapping ${\bf R= R}(\nu,s,t)$ at each fixed value
${\bf a}=(\nu,s)$ is a linear function of time with nonlinear oscillations,
which are described by the Landau-Lifshits equation (\ref{fer}) or its gauge
equivalent (\ref{schr}). The linear mean dependence reflects the fact that
the model under consideration is a model of free vortices.  The
collapse thus arises as a result of a running of one vortex into another.
Due to a continuous distribution of vortex lines,
their "density" increases infinitely in some point.

A similar situation takes place in the model  for large scale structure 
formation in
the Universe studied by  V.I.Arnold, Ya.B.Zeldovich
and S.F.Shandarin \cite{zeld}. In the basement of this model, the suggestion 
lies about
initial dust-like distribution of  masses, when their behavior can be
described by the zero-pressure hydrodynamic equations
\begin{eqnarray}
\rho_{t}+\mbox{div}\,\rho {\bf v}=0,  \\
\frac{d{\bf v}}{dt}={\bf v}_t+({\bf v\nabla)v}=0.
\end{eqnarray}
The integration of this system in the Lagrangian variables gives
that i) all fluid particles being free move with a constant velocity
\begin{equation}
\label{map2}
{\bf r=a+v(a})t,
\end{equation}
and ii) the density $\rho$ is expressed through the initial value and the
Jacobian of the mapping (\ref{map2}) as follows
\begin{equation}
\label{rho}
\rho({\bf r},t)=\frac{\rho_0({\bf a})}{J}.
\end{equation}
In the framework of this model, appearance of large scale structures
is connected with the breaking phenomenon resulting in  density singularities,
due to the Jacobian vanishing for the mapping (\ref{map2}). In a
typical situation, these structures have a pancake shape and can be
considered as pcurlo-galaxies. The formula, analogous to the Eq. (\ref{rho}),
takes place, as we have seen above, for the vorticity ${\bf \Omega}$ -- see
the Eq. (\ref{O_det}). However, there is a difference from Eq. (\ref{rho}),
connected with the vector nature of the ${\bf \Omega}$ field.

In the given task we will be interested in a geometric structure of a the
singularity at $t\to t_0$, but $t<t_0$, i.e., in some sense, at the initial
stage of  collapse, but not at the developed stage, which has the meaning for
astrophysics applications, but remains rather unclear for the incompressible
hydrodynamics, when  viscosity is necessary to be taken into account at 
small scales.

\subsection{Structure of the singularity}

Let us consider in more details the structure of the collapsing region in a
typical situation. First, it is clear from the above discussion that the
vorticity distribution  near singularity will be given by the
former expression (\ref{omega}). Second, the main features of the
singularity will be determined by the Jacobian, that is the denominator
of (\ref{omega}). Numerator ${\bf (\Omega_0(a)\nabla_a)R}$
( tangent vector to the lines) can be taken at the point ${\bf a=a_0}$,
$t=t_0$ and considered as a constant.

According to Eq.
(\ref{omega}) the Jacobian expansion contains positive definite
symmetric matrix $\gamma_{ij}$ of its second derivatives taken at the
point ${\bf a=a_0}$, $t=t_0$. At $t<t_0$, this matrix is assumed to be
non-degenerate: all its eigenvalues are positive, and the matrix itself
can be diagonalized. Hence it follows immediately
that compression along all principal axises in ${\bf a}$-space
will have  the same law: $l_a\sim \sqrt{t_0-t}$. Therefore, near the singular
point, vorticity ${\bf \Omega}$ will have the self-similar asymptotics
\begin{equation}
\label{omega1}
{\bf \Omega(r},t)=\frac{\Omega_0(\nu){\bf R}_s}{(t_0-t)(\alpha+
 \gamma_{ij}\eta_i\eta_j )},
\end{equation}
where ${\bf \eta}=\Delta{\bf a}/\sqrt{t_0-t}$ are  self-similar
variables in ${\bf a}$-space. However, the equation (\ref{omega1}) 
does not mean,
that compression in {\it  ${\bf r}$-space} will be the same.

When the Jacobian takes zero value, one of the eigenvalues of the 
Jacobi matrix becomes equal to zero.
This eigenvalue (denote it as $\lambda_1$ ), as it easily to see, coincides 
with the Jacobian (\ref{expansion})
at a small vicinity of the collapse point up to the almost constant multiplier
$\lambda_2\cdot\lambda_3$.

Represent now the Jacoby matrix $\hat J$ as a decomposition over
eigenvectors  of direct  ($\hat J|\psi^{(n)}>=\lambda_n|\psi^{(n)}>$)
and conjugated ($<\tilde \psi^{(n)}|\hat J=\lambda_n<\tilde \psi^{(n)}|$)
spectral problems
\begin{equation}
\label{exp}
J_{ik}\equiv \frac{\partial x_k}{\partial a_i}=
\sum_{n=1}^3 \lambda_n \psi^{(n)}_i
\tilde\psi^{(n)}_k.
\end{equation}
Here two sets of eigenvectors of  direct and conjugated problems are
mutually orthogonal
$$
<\tilde \psi^{(n)}|\psi^{(m)}>=\delta_{nm}.
$$
In a small vicinity of the collapse point eigenvalues $\lambda_{2,3}$ can
be considered as constants, while
$$
\lambda_1\equiv\frac{J}{\lambda_2\lambda_3}=(\lambda_2\lambda_3)^{-1}
[\alpha(t_0-t)+
\gamma_{ij} a_i a_j].
$$
Here, for simplicity, we have placed the origin at the point ${\bf a=a}_0$.
As for  the eigenvectors, they also can be considered constant.

Let us decompose the vectors  ${\bf x}$ and  ${\nabla_a}$ in Eq.
(\ref{exp}) through the corresponding bases, denoting their
appropriate projections as $X_n$ and $A_n$:
$$
X_n=<{\bf x}|\psi^{(n)}>, \,\,\, \frac{\partial}{\partial A_n}=
<\tilde\psi^{(n)}|\nabla_a>.
$$
In this case the vector ${\bf a}$ is written in terms of $A_n$  as follows
$$
a_{\alpha}=\sum_{n}\psi_{\alpha}^{(n)} | \tilde \psi^{(n)}|^2 A_n.
$$
As a result, Eq. (\ref{exp}) can be rewritten in the form
\begin{eqnarray}
\frac{\partial X_1}{\partial A_1}=\tau +\Gamma_{mn}A_mA_n,\\
\frac{\partial X_2}{\partial A_2}=\lambda_2, \qquad
\frac{\partial X_3}{\partial A_3}=\lambda_3 \label{system}.
\end{eqnarray}
Here, the matrix
$$
\Gamma_{mn}=\gamma_{\alpha \beta}(\lambda_2\lambda_3)^{-1}
\psi_{\alpha}^{(n)}\psi_{\beta}^{(m)}
|{\bf \tilde \psi^{(n)}}|^2|{\bf \tilde \psi^{(m)}}|^2 ,
$$
and parameter
$\tau=\alpha(t_0-t)(\lambda_2\lambda_3)^{-1}$
is assumed to be small.
It follows from here immediately that size reduction along the second
($X_2$) and the third ($X_3$) directions is the same as in the auxiliary
${\bf a}$-space, i.e.,  $\sim \sqrt{\tau}$, but along the "soft" direction
$X_1$ behaves like  $\tau^{3/2}$.
Respectively, in terms of new self-similar variables $\xi_1=X_1/\tau^{3/2}$,
$\xi_2=X_2/\tau^{1/2}$, $\xi_3=X_3/\tau^{1/2}$,
integration of the system  gives for $\xi_2$ and $\xi_3$ a linear
dependence on $\eta$ , while for $\xi_1$ it a cubic dependence
\begin{eqnarray}
\label{system1}
\xi_1=(1 +\Gamma_{ij}\eta_i\eta_j)\eta_1 +
\frac 1 2 \Gamma_{1i}\eta_i\eta_1^2+ \frac 13 \Gamma_{11}\eta_1^3,
\,\,\, i,j=2,3\\
\xi_2=\lambda_2 \eta_2, \,\,\,\xi_3=\lambda_3 \eta_3 \label{system2}.
\end{eqnarray}
Together with  (\ref{omega1}), the relations (\ref{system1}) and
(\ref{system2}) determine implicit dependence ${\bf \Omega(r},t)$.
The presence of two different self-similarities shows, that
the spatial vorticity distribution becomes strongly flattened in the
first direction, taking a pancake form as $t\to t_0$.

The direction of the field  ${\bf \Omega}$ can be  found from the
incompressibility condition $\mbox{div}\,{\bf \Omega}=0$. It is easy to see
that in the
leading order (as $t\to t_0$) the gradient of $J$ is determined by 
the soft direction ${\bf e_1}$:
$$
\nabla J\approx\tau^{-3/2}\frac{\partial J}{\partial \xi_1}{\bf e_1}.
$$
Contributions from  two other directions are small in the parameter $\tau$.

Hence, it follows that vector lines of the  field ${\bf \Omega}$  lie 
in the pancake plane that is in agreement with
transversality of motion of vortex lines (compare with Eq.(\ref{R})).

\section{Example of collapse in the degenerated case}

In the previous section we have considered collapse for a non-degenerated
situation, when the only one eigenvalue of the Jacoby matrix tends to zero at
the touching point. Now we shall consider an example of collapse, when
two eigenvalues of the Jacoby matrix vanish simultaneously at the
collapse point. We will examine an initial vorticity distribution with the
nontrivial topology of  vortex lines with linking number $N=1$.
This special distribution is the so-called  Hopf mapping. There are several 
ways how to construct the corresponding field ${\bf \Omega}$. 
We will keep here the approach of the paper \cite{KM}.

Following to that work, let us represent the field ${\bf\Omega}$ through the
${\bf n}$-field (${\bf n}^2=1$)
\begin{equation}
\label{n}
\Omega_{\alpha}({\bf r})=\frac{1}{32}\epsilon_{\alpha\beta\gamma}
{\bf(n\cdot[\partial_{\beta}n\times \partial_{\gamma}n])},
\end{equation}
where the ${\bf n}$-field is supposed to be a smooth function of coordinates tending to the constant
value ${\bf e}$ at the infinity $r\to\infty$.

It is easy to check that in accordance with Eq.(\ref{n}) each point ${\bf n =n}_0$
on the unit sphere ${\cal S}^2$ defines a closed vortex line .
Indeed, parameterization of the unit vector ${\bf n}$ through the spherical
angles $\theta$ and $\varphi$ allows to write down the field ${\bf\Omega}$
as follows
\begin{equation}
\label{n1}
{\bf \Omega}=\frac{1}{16}
[\nabla \varphi\times \nabla\cos\theta]),
\end{equation}
so that  the variables $\varphi$ and  $\cos\theta$ play the role of the Clebsch
variables. Thus, each vortex line coincides in this case with the
intersection of two surfaces $\varphi=const$ and $\cos\theta=const$, i.e.
a closed vortex line is the pcurlotype in ${\cal R}^3$  of a point on the
sphere ${\cal S}^2$. By the definition, the Hopf degree $N$ of mapping
${\cal R}^3\to {\cal S}^2$ is called an integer
number of linkages between arbitrary two vortex lines -- pcurlotypes of two
points on the unit sphere. The Hopf mapping (with $N=1$) is
given by the next relation:
\begin{equation}
\label{hopf}
{\bf (n\cdot\sigma)}=U{(\bf e\cdot\sigma})U^{\dagger},\,\,\,
U=\frac{1+i({\bf a\cdot\sigma})}{1-i({\bf a\cdot\sigma})},
\end{equation}
where ${\bf\sigma}$ are the Pauli matrices.

Expressing the vector ${\bf n}$ from here and substituting the result into
Eq.(\ref{n}), one can get (see \cite{Kamchatnov})
\begin{equation}
\label{omega0}
{\bf\Omega}_0({\bf a})=\frac{{\bf e}(1-a^2)+2{\bf a}({\bf e\cdot a})
+2[{\bf e}\times{\bf a}]}{(1+a^2)^3}.
\end{equation}
As it was shown in  \cite{Kamchatnov}, all the flux lines of this field
are circles, and each line is linked once with another line. By this
reason, the singularity formation is inevitable.

It is worth to note that the field (\ref{omega0})
has no singular points in the whole space
and its absolute value depends only on the absolute value
$|{\bf a}|$. The unit vector $\vec\tau$ is defined everywhere:
\begin{equation}\label{tau_a}
{\vec\tau}({\bf a})=\frac{{\bf e}(1-a^2)+2{\bf a}({\bf e\cdot a})
+2[{\bf e}\times{\bf a}]}{(1+a^2)}.
\end{equation}
The velocity of each ring is connected with the binormal ${\bf b}(\nu)$ and
the radius $r(\nu)$ by the relation
$$
{\bf V}(\nu)={\bf b}(\nu)/r(\nu).
$$
The radii $r(\nu)$  of rings and their orientations ${\bf b}(\nu)$ are
integrals of motion. Only  positions of centers of rings can change, and this
motion occurs with a constant velocity for each ring.

In the given problem, instead of the variables $\nu$ and $s$,
it is convenient to use directly the variables ${\bf a}$,
in which  the mapping ${\bf R}({\bf a},t)$ is written as
\begin{equation}\label{mapping}
{\bf R}({\bf a},t)={\bf a}+ t{\bf V}({\bf a}),
\end{equation}
where the velocity ${\bf V(a)}$ is expressed through the unit tangent vector
${\vec\tau}$ by means of the relation
\begin{equation}\label{V_a}
{\bf V}({\bf a})=
[{\vec\tau({\bf a})}\times(\vec\tau({\bf a})
\nabla_{\bf a})\vec\tau({\bf a})].
\end{equation}
Hence one can find the expression for the Jacobian
\begin{equation}\label{jacobian}
J({\bf a},t)=\mbox{det}\Big\|\hat I+
t\left(\frac{\partial{\bf V(a)}}{\partial{\bf a}}\right)\Big \|.
\end{equation}
It should be noted that the velocity of each ring is constant along the vortex line.
Therefore the determinant of the matrix
${\partial{\bf V(a)}}/{\partial{\bf a}}$,
which is the coefficient in front of $t^3$
in the previous expression, is equal to zero identically. As a result, $J$ has
a quadratic dependence on time $t$ only:
$$
J=1+tc_1({\bf a}) +t^2c_2({\bf a}).
$$
Singularity formation corresponds to the case $J\to 0$ at some point.
The collapse moment of time $t_0$  will be  determined by the coefficients
$c_1$ and $c_2$  so that $t_0$ will be a minimal positive root of the equation
$$
\mbox{min}_{\bf a}J({\bf a},t_0)\equiv J_{min}(t_0)=0.
$$
Calculations by means of Eq-s (\ref{tau_a}), (\ref{V_a}) and
(\ref{jacobian}) lead to the following expressions:
\begin{equation}
{\bf V}({\bf a})=\frac{-2}{(1+a^2)^2}
\left([{\bf e}\times{\bf a}](1-a^2)
+2{\bf a}({\bf e\cdot a})-2{\bf e}a^2\right),
\end{equation}
\begin{equation}\label{jacobianHopf}
J({\bf a},t)=\frac{(1+a^2)^3-8ta_3(1+a^2)+4t^2(1+a_3^2-a_1^2-a_2^2)}
{(1+a^2)^3},
\end{equation}
where $a_3$ is the projection of the vector ${\bf a}$ to the axis ${\bf e}$.

Analyzing the latter expression one can show that at $t<1$  a minimum of
the Jacobian is attained on the symmetry axis, i.e.,  at $a_1=0, a_2=0$.
In this case
\begin{equation}
J|_{axis}=\frac{(1-a_3^2)^2+4(t-a_3)^2}{(1+a_3^2)^2},
\end{equation}
hence it becomes clear that the singularity takes place at  $t_0=1,\,\,a_3=1$,
and the Jacobian tends to zero with the quadratic asymptotics.
Thus, in this example
$$
|\Omega|_{max}\sim (t_0-t)^{-2}.
$$
It can be easily verified also that the Jacoby matrix has two zero eigenvalues at
the touching point which eigenvectors lie in the
plane orthogonal to ${\bf e}$. In the collapse vicinity the field
${\bf \Omega}$ is directed along the vector ${\bf e}$.
Compression along this direction is linear in time $l_3\sim (t_0-t)$,
but in the perpendicular plane it is more fast, with the law
$\l_{1,2}\sim (t_0-t)^{3/2}$. As a result, the singularity structure
occurs strongly stretched  along the anisotropy axis.

\section*{Concluding remarks}

Considering the hydrodynamic model with the Hamiltonian ${\cal
H}=\int|{\bf\Omega}|d{\bf r}$, we have arrived at the conclusion
that each vortex line in the given system moves independently of
other lines. Just this property makes possible to form singularity
in a finite time for the generalized vorticity ${\bf\Omega(r},t)$
from smooth initial data. A typical singularity of this kind looks
like an infinite condensation of the vortex lines near some point.
Thus, the collapse in the integrable hydrodynamics has purely
inertial origin. If one will assume that this type of collapse is
possible also in the Euler hydrodynamics, then the asymptotics
of the vorticity near the singularity point (the point of vortex
"overturning")  will be given by
Eq-s (\ref{omega}) or (\ref{omega1}) in a non-degenerated situation.
That is already clear from
the general consideration. Namely, the curl of the velocity will
blow up as $(t_0-t)^{-1}$ . Exactly this dependence for vorticity
near a singular point has been observed in practically all
numerical simulations of the Euler equation, including the above
cited: \cite{PS}, \cite{kerr,kerr1}, \cite{GMG}, \cite{BP},
\cite{P}.

However, not all the simulations are regarded to
numerical integration of the Euler equations for continuous
distributions of vorticity. The first numerical experiments
\cite{PS} relate to investigation of collapse for two anti-parallel
vortex filaments which, as it was shown by Crow \cite{crow}, are
linearly unstable with respect to transverse perturbations ( about
the development of this direction see also \cite{P}). The theory of
collapse for two thin vortex filaments, as the nonlinear stage of
the Crow instability, was developed by V.E.Zakharov \cite{zakh},
\cite{zakh99} (see also \cite{KMD}). The conclusions of this theory
are in a good agreement with the numerical experiments up to the
distances comparable with a core size of filaments. This theory
predicts decreasing distance between vortex filaments as
$\sqrt{t_0-t}$. For smaller distances the cores of vortices loose
their round shape. They become flat, and the process of attraction
between vortices becomes more slow \cite{MH}, \cite{SMO}. The same
tendency was observed also in the numerical experiments of Kerr
\cite{kerr},  the most advanced (in  our opinion)
for the problem of reconnection, and where, unlike \cite{PS},
collapse of two anti-parallel vortices was simulated with
{\it continuously} distributed vorticity. Besides a natural
contraction of the minimal distance between the distributed
vortices, these simulations have shown, for the first time, the
formation of two singularities in two symmetric points.
While approaching a collapse moment in time, the explosive
growth of the maximal value of  vorticity  was observed with the law
$(t_0-t)^{-1}$. According to the recent publications of
Kerr \cite{kerr1}, the analysis of numerical data gave two
distinguished scales, one scale being contracted as the square root: $l_1\sim
(t_0-t)^{1/2}$, and another as the first power of time: $l_2\sim
t_0-t$.

In the work of Grauer, Marliani and Germaschewsky \cite{GMG},
the successfull attempt was undertaken to observe collapse for the initial 
condition not possessing a low symmetry.
The initial vorticity was concentrated in the vicinity of a cylinder, and
was modulated over the angle in such a way so that the simplest symmetries were
absent. In the present time this experiment has the best spatio-temporal
resolution. In this simulation appearance of a separate collapsing region was
observed with the vorticity growth at the center as  $(t_0-t)^{-1}$.

Thus, the results of the numerical simulations for ideal
hydrodynamics go along with the concept  of  collapse  as
a process of caustic formation for the solenoidal field.
There is an agreement for behavior of the vorticity
maximum. Concerning the spatial structure of the collapsing domain, we can make
only a qualitative agreement.
The results of the papers \cite{MH}, \cite{SMO} (where some
contraction of the vortex core was observed at the
initial stage of reconnection  neglecting viscosity) and
also the results of Kerr seem to support our theory.  We would like to pay
attention to the numerical results of the paper \cite{sulem} where for
short-time dynamics for the initial conditions in the form of the Taylor-Green
vortex and for random initial conditions it was observed formation of thin 
vortex layers with high vorticity that also support our predictions.

All these results allow one to say that the scenario
presented here looks like very plausible.

We would like to repeat once more, that if such scenario
takes place then  behavior of the vorticity near
of singular point is defined by Eq-s (\ref{omega}) or (\ref{omega1}).
The structure of this domain
should be highly anisotropic: in one of two directions perpendicular to the
vorticity, there is more rapid contraction ($\sim\tau^{3/2}$)
than with respect to other directions ($\sim \tau^{1/2}$).
The spatial distribution becomes close to the two-dimensional one familiar 
to a tangential discontinuity. The velocity of flow in this region can be 
approximated with a good
accuracy by the linear dependence
$$
v_{\perp}\sim \Omega_{max}X_1,
$$
i.e. the flow looks like a shear flow. The given type of collapse, according to the
classification of  \cite{ZK86}, belongs to a weak collapse:
the energy captured into singularity  
(with account of the viscosity $\nu$) tends to zero as $\nu\to 0$.
It is interesting to note that the dissipation rate
$\sim\int \Omega^2d{\bf r}$ from the
collapsing region also vanishes when  $\nu\rightarrow 0$.

One should note once more that unlike the model of free vortices considered
here, in the Euler hydrodynamics vortex lines interact pairly in
accordance with the Hamiltonian
\begin{equation}\label{HamEuleR}
{\cal H}_{Euler}=\frac{1}{8\pi}\int\!\int
\frac{{\bf R}_s(\nu,s)\cdot{\bf R}_{\xi}(\mu,\xi)}
{|{\bf R}(\nu,s)-{\bf R}(\mu,\xi)|}d^2\nu ds \,d^2\mu d\xi.
\end{equation}
{}From the viewpoint of investigation of the collapse problem it is very
important that the interaction function (i.e. the Green function for
the Laplace operator) has the singularity for equal arguments
${\bf R}(\nu,s)\to{\bf R}(\mu,\xi)$.
If this singularity would be absent, i.e. if the interaction function would be
completely regular, then {\it any} initial vortex line distribution
equivalent in the sense of iso-vorticity to a smooth field
${\bf\Omega}_0({\bf a})$ , even very singular distribution,
would produce a sufficiently smooth velocity field ${\bf v(r)}$.
In a smooth velocity field, an initial singularity of the generalized
vorticity could not disappear in subsequent moment of time, it could be only
transported and deformated by the fluid flow. As far as the equations of motion
for an inviscid substance are time-reversible, it follows from this, that the
formation of singularity from a smooth initial data also would not possible.
So, the existence and possible types of collapse of vortex lines, in the
systems with the quadratic on ${\bf\Omega}$ Hamiltonians,
depend on the asymptotics of the interaction function $G({\bf r}_1,{\bf r}_2)$
at ${\bf r}_1\to{\bf r}_2$. So, for a better understanding of the collapse
problem in hydrodynamics, it is reasonable to investigate such systems, for
which the interaction function has the asymptotics
$G\sim |{\bf r}_1-{\bf r}_2|^{-q}$ , and the exponent $q$
is not necessary equal to the unity.

What is the  influence of viscosity on the structure of collapsing region?
How does the given type of collapse have effect on turbulent spectra?
That are only few of the most important questions, which need investigation.
It would be interesting to verify numerically our hypothesis about
possibility of (quasi-)inertial collapse in ideal hydrodynamics, both in
Eulerian variables and in the vortex line representation.

\section*{Acknowledgments}

The discussions with V.E.Za\-kha\-rov, 
R.Z.Sag\-deev, V.V.Le\-bedev, G.E.Fal\-kovich,
A.Tsi\-no\-ber, I.Goldhirsch, N.Zabusky, J.Herring, R.Kerr, R.Pelz
of many questions concerned in the given article were very useful.
The authors are grateful to all of them.
E.K. thanks to P.L.Sulem for pointing out the paper \cite{sulem} to the 
attention. This work was supported by RFBR (grant 00-01-00929), by
Program of Support of the Leading Scientific Schools (grant 00-15-96007). V.R. 
thanks also the Fund of Landau Postdoc Scholarship
(KFA, Forschungszentrum, Juelich, Germany) for support. E.K. wishes to thank 
the Nice Observatory, where the paper was completed, for its hospitality and
financial support through the Landau-CNRS agreement.

\end{document}